\documentclass{raa_rb}           
\usepackage{graphicx,times}
\usepackage{natbib}
\usepackage{amssymb,amsmath}
\usepackage{epstopdf}
\usepackage{subcaption}
\bibpunct{(}{)}{;}{a}{}{,}

\usepackage[a4paper=true,driverfallback=dvipdfm,pagebackref=true]{hyperref}
\hypersetup{pdftitle = The title of my PDF, pdfauthor = My name, pdfsubject= The subject, pdfkeywords = keyword1 keyword2 keyword3} 
\hypersetup{colorlinks = true, linkcolor = green, anchorcolor = red, citecolor = blue, filecolor = red, pagecolor = red, urlcolor = red}

\begin{document}   
\title{Short-term H$\alpha$ line variations in Classical Be stars: 59 Cyg and OT Gem
}

\setcounter{page}{1}

   \author{Paul KT\inst{1,*}, Shruthi S Bhat\inst{1}, Annapurni Subramaniam\inst{2}
   }

   \institute{ Department of Physics, Christ University, Bangalore 560029, 
India;\\
        \and
             Indian Institute of Astrophysics, II Block, Kormangala, Bangalore 560034, India\\
             {* \it e-mail: 	paul.kt@christuniversity.in}
}

\abstract{We present the optical spectroscopic study of two Classical Be stars, 59 Cyg and OT Gem obtained over a period of few months in 2009. We detected a rare triple-peak H$\alpha$ emission phase in 59 Cyg and a rapid decrease in the emission strength of H$\alpha$ in OT Gem, which are used to understand their circumstellar disks. We find that 59 Cyg is likely to be rapid rotator, rotating at a fractional critical rotation of $\sim$ 0.80. The radius of the H$\alpha$ emission region for 59 Cyg is estimated to be  \( R_d/R_*\) $\sim$ 10.0, assuming a Keplerian disk, suggesting that it has a large disk.  We classify stars which have shown triple-peaks into two groups and find that the triple-peak emission in 59 Cyg is similar to $\zeta$ Tau. OT Gem is found to have a fractional critical rotation of $\sim$ 0.30, suggesting that it is either a slow rotator or viewed in low inclination. In OT Gem, we observed a large reduction in the radius of the H$\alpha$ emission region from $\sim$ 6.9 to $\sim$ 1.7 in a period of three months, along with the reduction in the emission strength. Our observations suggest that the disk is lost from outside to inside during this disk loss phase in OT Gem. 
}

   \maketitle


%

\section{Introduction}
\label{Sect1}


Classical Be stars, most commonly known as Be stars are very rapidly rotating main sequence B-type stars, which through a constrained process, form an outwardly diffusing gaseous, dust-free Keplerian disk \citep{rivinius2013}. This disk is formed from the material ejected from the fast-spinning central star. Be stars exhibit line emissions, mostly Balmer lines over the photospheric spectrum \citep{porter2003}. The emission lines originate from the geometrically thin, circumstellar disk rotating with near-Keplerian velocity surrounding the central star \citep{carciofi2006}. 

Be stars are well known variable stars and the period of spectroscopic variation, which includes short-term and long-term variations in emission lines, vary from few minutes to few decades \citep{porter2003}. The profile structure of the H$\alpha$ emission line vary from star to star which mainly depends on the inclination angle of the system as explained by \cite{struve1931}. \cite{catanzaro2013} used the classification scheme proposed by \cite{hanuschik1988} for the observed H$\alpha$ profile types as single peak, double peak, shell structure emission and also absorption. Spectroscopic monitoring of Be stars indicate H$\alpha$ emission line variations in equivalent width (EW), profile shape, V/R value \citep{dachs1987,hanuschik1988,hubert1994,hanuschik1996}. In this paper we discuss the H$\alpha$ line variability of two Classical Be stars i.e., 59 Cyg and OT Gem. The large variations observed in the H$\alpha$ emission line for these two stars are rapid and  in a short span of a few months.

Studying variation of emission lines is expected  to provide insights into the changes in the circumstellar disk. It can be used to derive the distribution of material and kinematics of the circumstellar disk \citep{shruthi2016}. In this paper, we present the emission line variability of 59 Cyg and OT Gem (see Table~\ref{Tab1}) based on spectroscopic data, 7 spectra of 59 Cyg and 8 spectra of OT Gem, obtained over a period of about three months in 2009. We discuss the observed features and changes in the spectra of these stars. We have used similar methods adopted by \cite{shruthi2016} to estimate the radius of the circumstellar disk using the H$\alpha$ line and also to determine the rotational velocity of the central star from prominent He{\sc i} absorption lines.

The paper is arranged as follows. The following section gives a brief overview of these two stars mainly on their spectroscopic variability from previous studies. Section 3 addresses the details of spectral observations and data reduction techniques. In section 4, we present the spectra and discuss the major results from the spectral line analysis of both the stars. The conclusions drawn from this study are listed in section 5.

\begin{table}[!htb]
\caption{Program stars}
\centering
\begin{tabular}{ccccccc}
\hline
\hline
$\bf HD$ & $\bf HR$ & $\bf Name$ & $\bf Spectral$ & $\bf RA$ & $\bf \delta$ & $\bf V$\\
 &  &  & $\bf Type$ &  \\ 
\hline
200120 & 8047 & 59 Cyg & B1 Ve & 20 59 49.55716 & +47 31 15.4216 & 4.75\\
58050 & 2817 & OT Gem & B2 Ve & 07 24 27.64809 & +15 31 01.9061 & 6.41\\
\hline
\end{tabular}
\label{Tab1}
\end{table}

\section{Previous studies}

\subsection{59 Cyg}
\label{Sect2.1}

59 Cyg is a well known Be star because of its pronounced spectral variations in emission line profiles and intensities which is summarized by \cite{barker1982} and \citep{harmanec2002}. It was first discovered of its emission lines in 1904 by Cannon and had shown emission components with variable intensities all throughout except in 1912 and 1916 \citep{hubert1981,barker1982}. Strong V/R variations were reported in 1926-1929, 1941-1942 and in 1946-1948; a slow increase of emission was also observed from 1945-1950 by \cite{merrill1949}. Afterwards, a quiescent phase had set in from 1953-1970 with the emission features being relatively stable \citep{hubert1981}. Emission was at maximum in 1956 and in 1961 and at minimum in 1967 \citep{moujtahid1998}

	\cite{kogure1982} mentioned that the star showed the first gradual strengthening of emission lines in 1971-1972 and came up with rich shell lines in 1973 June. This constituted the first shell phase, in which strong shell absorption lines were detected in Balmer lines (upto H30), He{\sc i}, Mg{\sc ii} and in some singly ionized metals \citep{doazan1975,hubert1981}. In the period 1973-1974, shell lines changed to second strengthening of emissions with asymmetric profile (V/R $>$ 1) and H$\beta$ as single emission line with a very deep core. With the declining of emission components and the appearance of double emission peaks, 59 Cyg proceeded to the second shell phase in 1974 October, which lasted till 1975 March and was stronger than the first shell phase \citep{hubert1981}. \cite{barker1982} describes the 1974-1975 shell phase as a 160-day episode with a general outline of spectral changes and also the detailed line profile variations.

	Soon after, in 1978, a new Be phase began to develop which was observed in the far UV and optical regions by \cite{doazan1989}. They observed the star again in the time interval 1978-1987 and saw the V/R variability and intensity changes in the emission of H$\alpha$. \cite{doazan1985} gives the V/R variability of H$\alpha$ on a long term to be about 2 years. \cite{barker1983} describes the transient emission events of the star, after the minimum of 1977, with a V/R variation on a short-lived quasi-period no longer than 28 days. 

	59 Cyg is a part of the multiple system (ADS 14526) of the Trapezium type. Optical components are separated from 59 Cyg A by \(20^{\prime\prime}\), \(26^{\prime\prime}\) and \(38^{\prime\prime}\) respectively. It was  confirmed as a single-lined spectroscopic binary by studying the variations of photospheric lines, like He{\sc i} 4471 with a period of 28.1702 $\pm$ 0.0014 day by \cite{rivinius2000} who compared it to $\phi$ Per. 59 Cyg was confirmed to be a Be + sdO binary as the companion was suspected to be a compact object by \citep{maintz2005}.

\subsection{OT Gem}
\label{Sect2.2}

OT Gem has exhibited strong spectral variations during the past years. The first evidence of the emission changes was presented by \cite{merrill1943}. The spectroscopic behaviour between 1954-1975 was described by \cite{hubert1979}, where the strong emission of Balmer and Fe{\sc i} lines was clearly seen. The strength of the emission steadily decreased after the maximum in 1961-1962, with slight variations and finally reaching a minimum at the end of 1980 \citep{hubert1982}. \cite{dachs1986} described the H$\alpha$ emission as a single, sharp emission peak with slightly decreasing strength from 1981 to 1983 which is in good agreement with the H$\alpha$ measurement obtained in 1981 by \cite{andrillat1983}. This slight decrease of Balmer emission line intensity is from 1961 as described by \citep{hubert1982} which had still continued to 1981-1983.

	\cite{hanuschik1996} classified OT Gem as a non-shell star and concluded that a significant part of the disk is projected against the sky. \cite{bozic1999} compared the variations of OT Gem to $\omega$ CMa and said that the physical process might be same for the two stars. The H$\alpha$ peak intensities of 4.3 \citep{andrillat1983} and 3.2 \citep{doazan1991} were reported. \cite{poretti1982} classified OT Gem as a $\gamma$ Cas type but \cite{ferro1998} considered it to be a mild-$\gamma$ Cas because the intensity of the emission at H$\alpha$ reached only 3 in OT Gem unlike 5 in $\gamma$ Cas. They also mention about the scarcity of available spectroscopic data for this star to give any correlation with the photometric measurements. \cite{catanzaro2013} reports a triple-peaked structure for H$\alpha$ in one of the nights between 2008 and 2009.

\section{Observations and data reduction}
\label{Sect3}

\begin{table}[!htb]
\centering
\caption{Journal of observations for 59 Cyg and OT Gem}
\begin{tabular}{cccc}
\hline
\hline
$\bf Star$ & $\bf Date\ of$ & $\bf Spectral$ & $\bf No.\ of\ spectra$ \\
$\bf Name$ & $\bf Observation$ & $\bf Range$ & $\bf (Integration\ time$ \\
 & $\bf in\ 2009$ & $\bf $\AA$ $ & $\bf in\ seconds)$ \\
\hline
59 Cyg & 1 June & 3800 -- 4300 & 1 (1800)\\
 & 2 June & 3800 -- 4300 & 1 (1800)\\
 & 29 June & 3800 -- 4300 & 2 (1800)\\
 & 11 July & 6200 -- 6800 & 1 (2700)\\
 & 23 July & 6200 -- 6800 & 2 (2400)\\
 \hline
OT Gem & 4 February & 6200 -- 6800 & 1 (2700) \\
 & 27 April & 6200 -- 6800 & 2 (1800, 2700)\\
 & 28 April & 6200 -- 6800 & 2 (1800)\\
 & 29 April & 6200 -- 6800 & 1 (2700)\\
 & 30 April & 6200 -- 6800 & 1 (2400)\\ 
 & 1 May & 6200 -- 6800 & 1 (2400)\\ 
\hline
\label{Tab2}
\end{tabular}
\end{table}

%

The spectra of 59 Cyg and OT Gem were acquired during several observation runs from February 2009 to July 2009 and the journal of observations is given in Table~\ref{Tab2}. The spectra were obtained using the Universal Astronomical Grating Spectrograph (UAGS) at the Cassegrain focus of the 1.0m Carl Zeiss reflector located at Vainu Bappu Observatory, Kavalur, India which is operated by the Indian Institute of Astrophysics (IIA). The CCD consists of \(1024 \times 1024\) pixels of 24 $\mu$m size, where the central \(1024 \times 300\) pixels were used for spectroscopy. The typical readout noise is of about \(4.8e^-\) and the gain is \(1.22e^-/\)ADU. Bausch and Lomb 1800 lines per millimetre grating was used, which in combination with the slit provided a resolution of 1~\AA ~at H$\alpha$. The medium resolution data taken in the wavelength region 3800 -- 4600~\AA ~included absorption lines like H$\gamma$ to H$\theta$ and also He{\sc i} lines and data taken in the range 6200 -- 6800~\AA ~had H$\alpha$ in emission.

The reduction of all the spectra was performed using several routines in the NOAO/IRAF~\footnote{IRAF is distributed by the National Optical Astronomy Observatory, which is operated by the Association of Universities for Research in Astronomy, Inc.}(Image Reduction and Analysis Facility) package. The wavelength calibration was performed using Fe-Ar arc lamp spectra. Typical S/N near H$\alpha$ for 59 Cyg is $\sim$200 from 3 spectra and for OT Gem is $\sim$100 from 8 spectra.

The spectra were initially normalized to the continuum. IRAF tasks were later used to measure parameters of the emission line profiles,  such as Equivalent Widths (EW), Full Width at Half Maximum (FWHM), V/R ratios, peak separations ($\Delta$V) and \( I_p/I_c\). All the measurements are reported in the next section.


\section{Analysis and discussion}
\label{Sect4}
\begin{figure}[''h'']
    \begin{subfigure}{0.5\textwidth}
    \centering
    \includegraphics[width=6.5cm,  height=6.5cm]{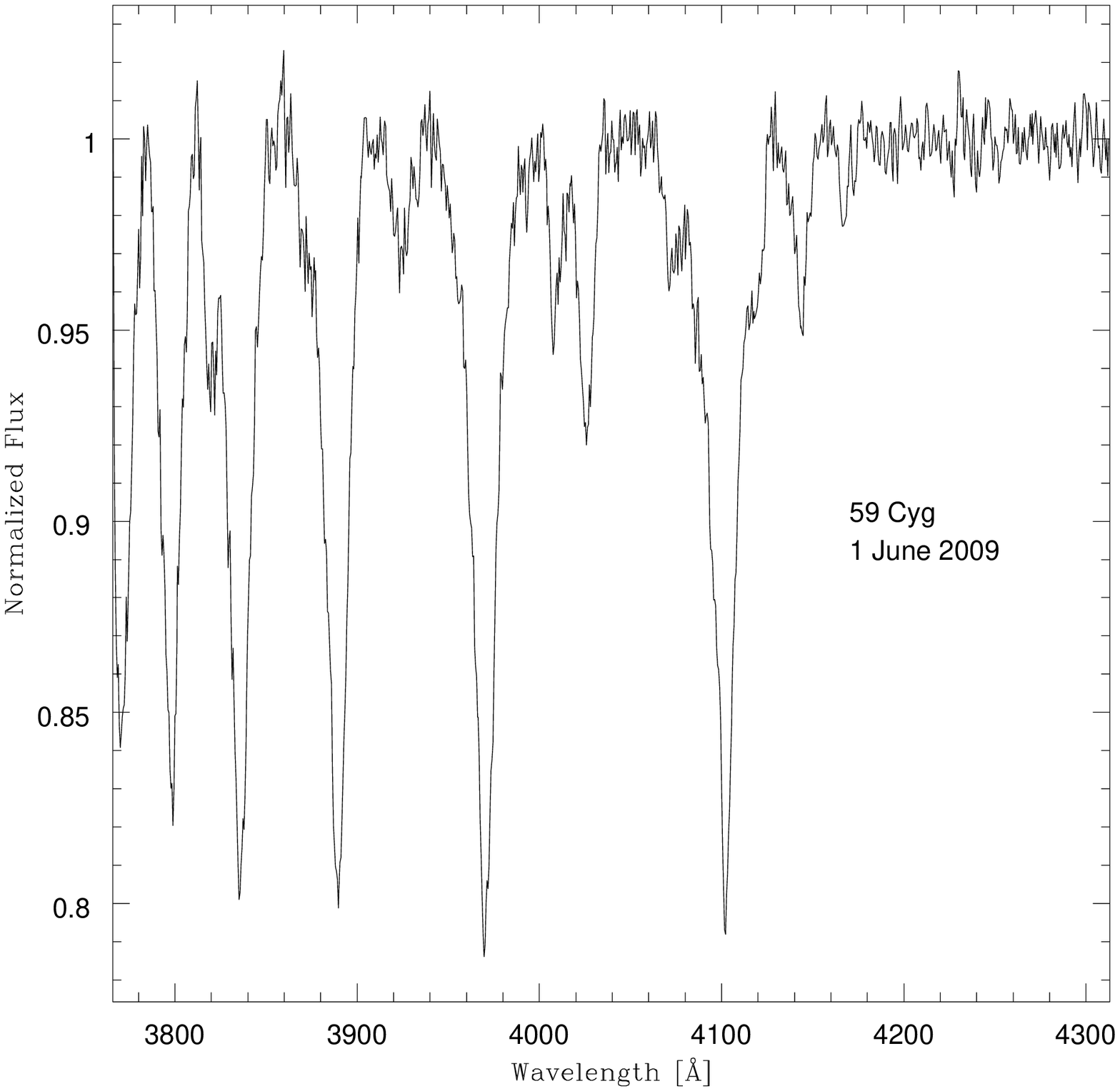}
    \end{subfigure}%
    \hspace{0.5cm}
    \begin{subfigure}{0.5\textwidth}
    \centering
    \includegraphics[width=6.5cm,  height=6.5cm]{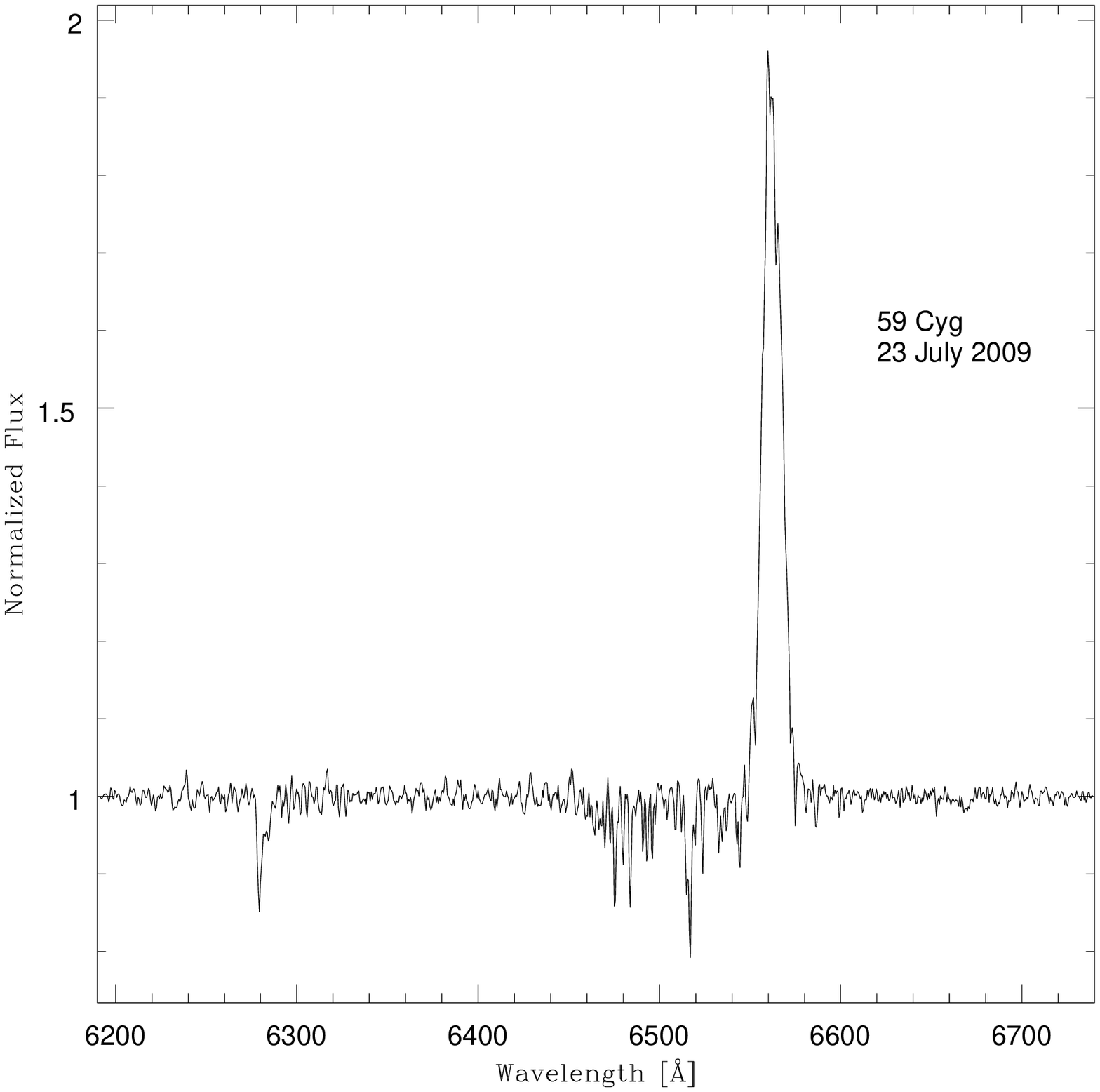}
    \end{subfigure}%
	\caption{Representative sample spectra of 59 Cyg (Left: 3800 -- 4300~\AA ~showing H$\delta$, H$\epsilon$, H$\zeta$, H$\eta$, H$\theta$ along with He{\sc i} 4026~\AA ~and Fe{\sc ii} absorption lines, Right: 6200 -- 6800~\AA ~showing H$\alpha$ emission along with He{\sc i} 6678~\AA and Fe{\sc ii} absorption lines)}
	\label{Fig1}
\end{figure}

59 Cyg was observed in June 2009 in the shorter wavelength region to investigate the He{\sc i} lines. It was also observed in July 2009 in the H$\alpha$ region.  Representative sample spectra for 59 Cyg in different wavelength regions is shown in Figure \ref{Fig1}. OT Gem was observed from February to May, 2009 in the H$\alpha$ region only. In the following section, we discuss the rotational velocity of the two stars. Sect. \ref{Sect4.2} deals with variability of the H$\alpha$ emission line of both the stars.

\subsection{Rotational velocity estimation}
\label{Sect4.1}

Be stars belong to the most rapidly rotating class of non-degenerate stars and it is also observed that a few stars may be rotating very close to the critical velocity \citep{rivinius2013}. In this study, to estimate the rotational velocity ($\textit{v}$ sin $\textit{i}$), we have considered the He{\sc i} absorption lines in the blue spectral region, He{\sc i} $\lambda$4009, $\lambda$4026 and $\lambda$4143 (refer left panel of Figure \ref{Fig1}), for the star 59 Cyg. These lines are assumed to be unaffected by the emission from the disk. 

\cite{steele1999} derived the rotational velocities for a sample of 58 Be stars and made a fit to the FWHM - $\textit{v}$ sin $\textit{i}$ correlation of \cite{slettebak1975}. They obtained the relations Eq. 1 - 4  in their paper for four different He{\sc i} lines. We estimate the $\textit{v}$ sin $\textit{i}$ only for He{\sc i} $\lambda$4026 and $\lambda$4143 using the respective relations between FWHM and $\textit{v}$ sin $\textit{i}$ as given in their paper. We do not have a similar relation given for He{\sc i} $\lambda$4009 in \cite{steele1999}, so we use a basic relation between the FWHM of the He{\sc i} line and the $\textit{v}$ sin $\textit{i}$ of the star as given in the relation below. 
\begin{equation}
vsin\emph{i} = \frac{c(FWHM)}{2\lambda\sqrt{ln2}} 
\label{eq1}
\end{equation}

The FWHM of the selected He{\sc i} lines and the estimated $\textit{v}$ sin $\textit{i}$ values are shown in Table~\ref{Tab4} for 59 Cyg. It can be seen that the values derived using He{\sc i} $\lambda$4009, are similar to those from the other two He{\sc i} lines. The average $\textit{v}$ sin $\textit{i}$ estimated for this star from 4 spectra is shown in Table~\ref{Tab5}. The error tabulated for all the values corresponds to the standard deviation. The average values obtained were compared with the values from \cite{rivinius2006}, \cite{harmanec2002} and \cite{slettebak1982}. He{\sc i} $\lambda$4026 was well resolved in most of the spectra and was consistently detected well throughout the sample, and the estimated $\textit{v}$ sin $\textit{i}$ is found to be within the error. All the individual averages of He{\sc i} lines, as well as the collective average calculated from the individual averages for 59 Cyg matches with the literature values, except that of \cite{slettebak1982}. 
 
\begin{table}[!htb]
\centering
\caption{FWHM and \emph{v}\ sin\ \emph{i} measurements for 59 Cyg from the spectra using He{\sc i} 4009, 4026 and 4143\AA}
\begin{tabular}{ccccccc}
\hline
\hline
$\bf Date\ of$ & \multicolumn{2}{c}{$\bf HeI\ 4009 $\AA$ $} & \multicolumn{2}{c}{$\bf HeI\ 4026 $\AA$ $} & \multicolumn{2}{c}{$\bf HeI\ 4143 $\AA$ $}\\
$\bf Observation$ & $\bf FWHM\ ($\AA$)$ & $\textit{v}$ sin $\textit{i}$ $(\rm kms^{-1})$ & $\bf FWHM\ ($\AA$)$ & $\textit{v}$ sin $\textit{i}$ $(\rm kms^{-1})$ & $\bf FWHM\ ($\AA$)$ & $\textit{v}$ sin $\textit{i}$ $(\rm kms^{-1})$\\
\hline
01/06/09	 & 7.9 & 355.5 & 8.5 & 389.5 & 8.7 & 387.2\\
02/06/09 & 11.3 & 506.9 & 7.9 & 362.0 & 8.3	& 369.4\\
29/06/09 & 9.3 & 415.6 & 6.9 & 316.2 & 10.7 & 476.3\\
29/06/09 & 10.1 & 453.0 & 8.5 & 389.5 & 10.1 & 449.6\\
\hline
\end{tabular}
\label{Tab4}
\end{table}

\begin{table}[!htb]
\centering
\caption{Rotational velocity parameters. \emph{v}\ sin\ \emph{i} averaged from 4 spectra for the He{\sc i} lines for 59 Cyg is compared with two other estimations; $\omega$ was calculated using \(v_c\) given in Table 2 of \cite{yudin2001}, who interpolated values given by \cite{moujtahid1999}.}
\begin{tabular}{ccc}
\hline
\hline
$\bf Parameters$ & $\bf Reference$ & $\bf Value$\\
\hline
$\textit{v}$ sin $\textit{i}$ & 4009~\AA & 433 $\pm$ 32\\
$(\rm kms^{-1})$ & 4026~\AA & 364 $\pm$ 17\\
							&  4143~\AA & 421 $\pm$ 25\\
							& Average &  406 $\pm$ 25\\
							&  \cite{rivinius2006} & $\geq$379\\
							& \cite{harmanec2002} & 450\\
                             &  \cite{slettebak1982} & 260\\                          
\hline                              
$ v_c\ (\rm kms^{-1}) $ & Yudin & 520\\       
$\omega$ &  4009~\AA & 0.83\\
& 4026~\AA & 0.7\\
& 4143~\AA & 0.81\\
& Average & 0.78\\
\hline 
\end{tabular}
\label{Tab5}
\end{table}

\cite{harmanec2002} studied the long term as well as rapid variability of 59 Cyg using photometry and spectroscopy. They estimated the basic physical properties of 59 Cyg. \cite{rivinius2006} categorized 59 Cyg into a different class which have emission $\Leftrightarrow$ emission \& shell transitions. These type of stars show a very high $\textit{v}$ sin $\textit{i}$ values, which matches with our estimation.  

The critical velocity, \( v_c \) value was taken from \cite{yudin2001} which was estimated for a particular spectral class and luminosity class. Thus, for the spectral type indicated in Table~\ref{Tab1}, \( v_c \) was obtained and critical fractional rotation, $\omega$ given by \( v\ sin\ \emph{i}/{v_c} \) was estimated. The uncertainty in $\omega$ is not only from $\textit{v}$ sin $\textit{i}$ but also from the spectral and the luminosity classes \citep{rivinius2006}. The critical fractional rotation obtained in our study using the He{\sc i} lines was compared with the values obtained by \citep{rivinius2006}. $\omega$ value estimated using He{\sc i} 4026~\AA ~seem to be very close to the literature value, but those estimated using the 4009 and 4143~\AA ~ lines show about 10\% deviation. This is due to the fact that the $\textit{v}$ sin $\textit{i}$ itself is different for the two He{\sc i} lines. 

\begin{table}[!htb]
\centering
\caption{FWHM and \emph{v}\ sin\ \emph{i} measurements for OT Gem from the spectra using He{\sc i} 6678\AA}
\begin{tabular}{ccc}
\hline
\hline
$\bf Date\ of$ & $\bf FWHM$ & $\textit{v}$ sin $\textit{i}$\\
$\bf Observation$ & $\bf (\AA) $ & $(\rm kms^{-1})$\\
\hline
04/02/09 & 3.9 & 105.5\\
27/04/09 & 4.6 & 125.2\\
27/04/09 & 5.0 & 135.3\\
28/04/09 & 4.5 & 120.7\\
28/04/09 & 4.5 & 122.4\\
29/04/09 & 5.0 & 134.0\\
30/04/09 & 5.0 & 133.5\\
01/05/09 & 4.8 & 129.3\\
\hline
\end{tabular}
\label{Tab6}
\end{table}

For OT Gem, though there were no spectra available in the blue region, He{\sc i} 6678 \AA ~was available in the red region spectra. This line is usually affected by the emission from the disk. We can consider the line to be minimally affected during our observation since the star is in a disk loss phase and hence the emission is declining in its strength. Measurements for OT Gem like that of 59 Cyg was obtained and $\textit{v}$ sin $\textit{i}$ was estimated using Eq. \ref{eq1} and is shown in Table~\ref{Tab6}. The average $\textit{v}$ sin $\textit{i}$ from 8 He{\sc i} profiles for OT Gem is found to be 126 $\pm$ 4 $\rm kms^{-1}$. \cite{bozic1999} quoted a value of 130 $\rm kms^{-1}$. Our value matches very well with that of the literature. The critical velocity, \( v_c \) = 483 $\rm kms^{-1}$ value was taken from \cite{yudin2001} based on the spectral type of OT Gem. Thus the critical fractional rotation for OT Gem is found to be only $\sim$ 0.26. This indicates that the star OT Gem is very slow rotator and this might be an effect also due to the low inclination angle. 

\subsection{Variability of the H$\alpha$ line}
\label{Sect4.2}

Circumstellar decretion disk of Be stars are formed due to the material ejected from the central star and is equatorially flattened. The emission lines seen in the spectra arise from the disk and the most common emission seen for Be stars is the H$\alpha$ emission line. The radius of the H$\alpha$ emission disk is estimated according to the rotational velocity law as shown in Eq. \ref{eq2} for both the stars in the following subsections. The extent of the H$\alpha$ emission region, \( R_d\) is estimated in terms of the stellar radius \( R_*\) by assuming the region to be in Keplerian orbit around the star \citep{huang1972}. \( R_d\) is also estimated for non-Keplerian orbit by changing the rotational parameter \textit{j} from 1/2 to 1. 
\begin{equation}
\frac{R_d}{R_*} = \left(\frac{2\ v\ sin\ \emph{i}}{\Delta V}\right)^\frac{1}{j} 
\label{eq2} 
\end{equation}

\begin{figure}[!htb]
    \centering
    \includegraphics[width=10cm,  height=8.5cm]{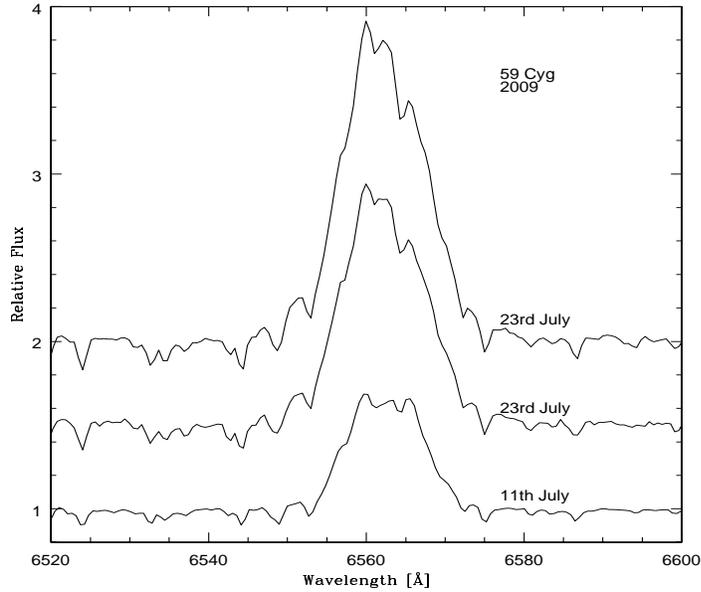}
    \caption{Time Series of 59 Cyg H$\alpha$ line observed in July 2009; (Spectra are offset and labelled with the observation date, the oldest appears at the bottom and most recent at the top. Note that
although the spectra are displayed evenly spaced, they are not evenly distributed in time.)}
\label{Fig2}
\end{figure}

\begin{figure}[!htb]
    \centering
    \includegraphics[width=10cm,  height=8.5cm]{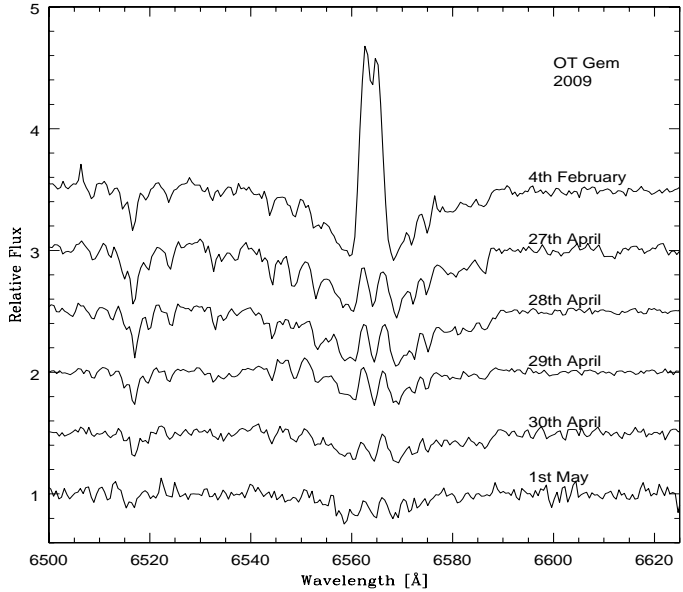}
    \caption{Time Series of OT Gem H$\alpha$ line from February to May 2009; (Spectra are offset and labelled with the observation date, the oldest appears at the top and most recent at the bottom. Note that
although the spectra are displayed evenly spaced, they are not evenly distributed in time.)}
\label{Fig3}
\end{figure}

The variability of the H$\alpha$ emission line is seen for the two stars and are discussed separately in the following subsections. The time series of the H$\alpha$ profile of 59 Cyg is shown in Figure \ref{Fig2} and OT Gem in Figure \ref{Fig3}.

\subsubsection{59 Cyg - triple peak in H$\alpha$ emission}

59 Cyg was observed once on 11 July and twice on 23 July 2009 in the H$\alpha$ region. Triple-peak feature in the H$\alpha$ emission was observed on all the nights, where the profile shows three emission peaks in the H$\alpha$ spectral line. The third central peak became prominent on 23 July compared to 11 July and appeared to be more connected to the V peak of the profile. V/R ratio was measured considering the extreme peaks and the value was initially 1 and then changed to $\sim$ 1.3 on 23 July. The \(I_p/I_c\), EW, $\Delta$ V and V/R of H$\alpha$ line for 59 Cyg has been shown in Table ~\ref{Tab7} for the two observed dates. The average values and the estimation of the radius of the disk is shown in Table ~\ref{Tab8}. The error in EW and $\Delta$V are the standard deviation of the available observations.

\begin{table}[!htb]
\centering
\caption{Measurements of parameters using H$\alpha$ emission for 59 Cyg}
\begin{tabular}{ccccc}
\hline
\hline
$\bf Date\ of$ & \(I_p/I_c\) & $\bf EW$ & $\Delta$ $\bf V$ & $\bf V/R$\\
$\bf Observation$ & & $\bf (\AA) $ & $(\rm kms^{-1})$ & \\
\hline
11/07/2009 & 1.69 & -11.0 & 259.8 & 1.04\\
23/07/2009 & 1.96 & -13.2 & 243.3 & 1.3\\
23/07/2009 & 1.96 & -12.7 & 241.4 & 1.33\\
\hline
\end{tabular}
\label{Tab7}
\end{table}

\begin{table}[!htb]
\centering
\caption{H$\alpha$ emission line parameters and estimation of the radius of the disk for 59 Cyg}
\begin{tabular}{cccccc}
\hline
\hline
\(I_p/I_c\) & $\bf EW$ & $\Delta$ $\bf V$ & $\bf \textit{v}\ sin\ \textit{i}$ \textsuperscript{\dag} & \multicolumn{2}{c}{ $\bf R$\begin{scriptsize}d\end{scriptsize}/$\bf R$\begin{scriptsize}*
\end{scriptsize}}\\
& $\bf (\AA) $ & $(\rm kms^{-1})$ & $(\rm kms^{-1})$ & $\textit{j} = 1/2$ & $\textit{j} = 1$\\
\hline
 & & & 364 & 8.63 & 2.94\\[0ex]
 & & & 421 & 11.55 & 3.40\\[1ex]
\raisebox{2ex}{1.87} & \raisebox{2ex}{-12.26 $\pm$ 0.66} & \raisebox{2ex}{248.2 $\pm$ 5.9} & 379 & 9.33 & 3.05\\[0ex]
 & & & 260 & 4.39 & 2.10\\[1ex]
\hline
\multicolumn{6}{l}{\textsuperscript{\dag}\footnotesize{Refer Table~\ref{Tab5}}}\\
\end{tabular}
\label{Tab8}
\end{table}

\cite{slettebak1992} assumed Keplerian geometry for the circumstellar disk and estimated the range for the radius of the H$\alpha$ emitting region in classical Be stars to be 7 -- 19 \( R_*\). Comparing our estimation i.e., $\sim$ 10 \( R_*\) to this range, we can conclude that 59 Cyg has a circumstellar disk, whose extent is similar to those seen for many other Be stars. We have not given the errors in the values of the radius as it would suffice to give only a typical range of them. The range of the radius of the disk is given by 8.6 -- 11.6 \( R_*\). An inspection of the H$\alpha$ profile suggests that the emission is going through a V/R variation and that it is shifting from  \(V \sim R\) to \(V > R\), see Table ~\ref{Tab7}. We see that the the EW increased during the period mainly due to the increased emission in the violet part of the profile. All the three spectra show the presence of the peak in between the V and R peaks, giving rise to three peaks. We checked the BeSS database for spectra taken immediately after our observations, we found spectra obtained on 17 August 2009, 6, 7 and 10  September 2009. An inspection of the spectra obtained on 17th August 2009 suggest that, the profile is dominated by emission in the V, and the emission in the R side is found to be low, suggesting a V-dominated profile. The spectra obtained in September 2009, show that the R side of the profile is strengthened almost to the level of V, but still with a profile V/R slightly greater 1.0.

\subsubsection{OT Gem - Rapid disk loss phase}

OT Gem was observed only in the H$\alpha$ wavelength region during our observations. It was observed totally 8 times on 6 nights. This star showed a very good strength in H$\alpha$ emission in February but later emission decreased in strength and went below the continuum level. Even though the emission had reduced significantly, the two peaks were still seen clearly in the spectra obtained in April and May. The \(I_p/I_c\), EW, V/R, $\Delta$ V and \( R_d/R_*\) of H$\alpha$ line for OT Gem are tabulated in Table ~\ref{Tab9} for all the observation dates. \( R_d/R_*\) was estimated using Eq. \ref{eq2}, by considering the $\textit{v}$ sin $\textit{i}$ to be 126 $\rm kms^{-1}$ for OT Gem as discussed in Sect. \ref{Sect4.1}. Table ~\ref{Tab9} clearly shows the change in the strength of the H$\alpha$ emission line from February to April. The measurements of \(I_p/I_c\), EW and $\Delta$ V, all show a significant change in strength. 
\begin{table}[!htb]
\centering
\caption{H$\alpha$ emission line parameters and estimation of the radius of the disk for OT Gem}
\begin{tabular}{ccccccc}
\hline
\hline
$\bf Date\ of$ & \(I_p/I_c\) & $\bf EW$ & $\bf V/R$ & $\Delta$ $\bf V$ & \multicolumn{2}{c}{ $\bf R$\begin{scriptsize}d\end{scriptsize}/$\bf R$\begin{scriptsize}*
\end{scriptsize}}\\
$\bf Observation$ & & $\bf (\AA) $ &  & $(\rm kms^{-1})$ & $\textit{j} = 1/2$ & $\textit{j} = 1$\\
\hline
04/02/09 & 1.34 & -4.1 & 1.09 & 96.0  & 6.89 & 2.63\\
27/04/09 & 0.95 & -0.7 & 0.83 & 181.8 & 1.92 & 1.39\\
27/04/09 & 0.96 & -0.7 & 0.92 & 165.7 & 2.31 & 1.52\\
28/04/09 & 0.96 & -0.8 & 1.00 & 194.8 & 1.67 & 1.29\\
28/04/09 & 0.97 & -0.7 & 1.12 & 191.2 & 1.74 & 1.32\\
29/04/09 & 0.95 & -0.8 & 0.84 & 174.6 & 2.08 & 1.44\\
30/04/09 & 0.96 & -0.6 & 2.11 & 190.3 & 1.75 & 1.32\\
01/05/09	 & 0.97 & -0.8 & 1.98 & 167.7 & 2.26 & 1.50\\
\hline
\end{tabular}
\label{Tab9}
\end{table}

The disk of OT Gem is seen to be dissipating and the observed variability is a result of the long-term variation associated with the disk i.e., the disk loss phase of a Be star. The period of such long-term variability for Be-stars can be generally from several years to several decades \citep{rivinius2013}. OT Gem was observed by few amateur astronomers and the spectra is archived in BeSS database \citep{neiner2011}. From BeSS, it is observed that in 2009, during the time of our observation, there is only one spectra in March and as expected, it shows a decrease in strength compared to our February observations. Later, the star continued to have a decreased strength in 2010 but suddenly had an increase in emission strength during 2011. It again decreased in strength during 2012, eventually going into absorption in 2013. The ``B-phase" has continued till March 2016 and is showing an emission again from April 2016. The period of our observation might be during short rapid phase within a period of slowly dissipation. The recent rebuilding of the disk can be an interesting observation and continuous monitoring of this star is being carried out. 

As previously mentioned, the radius of the H$\alpha$ emitting region given by \cite{slettebak1992} is  7 -- 19 \( R_*\). The range of radius during our observations for OT Gem is between 1.7 -- 6.9 \( R_*\). OT Gem at first in February has a radius which is already below the general observed range and later in April, it goes below the limit. In this episode of disk loss, the outer disk is lost and the feeble emission comes from only the inner disk. OT Gem lost about a little more than half of its disk within three months. We have captured this short-term phenomenon of sudden loss of disk and detect a change in the disk radius as well. This clearly indicates the need of continuous monitoring of such systems. There are very less spectroscopic study of OT Gem in the past. Thus our study would add significantly for any future variability study of this star. 

\subsection{Discussion}
\label{Sect4.3}

In this study, we have captured two different type of short-term variation in Classical Be stars, helpful in providing insights into the properties of their circumstellar disk. We also estimated various parameters of the two stars, 59 Cyg and OT Gem. We estimated the rotational velocity, $\textit{v}$ sin $\textit{i}$ of the stars to be 406 $\pm$ 25$\rm km $ $ s^{-1}$ for 59 Cyg and 126 $\pm$ 4$\rm km $ $ s^{-1}$ for OT Gem. 59 Cyg has been classified as emission $\Leftrightarrow$ emission \& shell transition type of star by \cite{rivinius2006}. Thus, the very high rotational velocity generally seen for shell stars is observed even for 59 Cyg. Whereas, OT Gem has been classified as a non-shell star by \cite{hanuschik1996} and thus the estimated low rotational velocity is consistent with the definition of shell star being an edge-on star and non-shell star having a lower inclination angle. By assuming the critical velocity based on the spectral class, we estimated that the fractional critical rotation is about 0.78 for 59 Cyg, suggesting that the star is rotating very close to the break-up velocity. For OT Gem, the fractional critical rotation is only about 0.26 which indicates that it is a very slow rotator, or viewed in low inclination. We also estimated the radius of H$\alpha$ emission, and is found to be in the range, 8.6 -- 11.6 \( R_*\) for 59 Cyg and 1.7 -- 6.9 \( R_*\) for OT Gem. In summary, we find that 59 Cyg is a fast rotator with the H$\alpha$ emission region far from the star and OT Gem is a slow rotator or viewed in low inclination, with a feeble emission coming from the inner disk very close to the star. 

59 Cyg is known to be a Be binary with a sdO companion \citep{maintz2005}. It is often compared to $\phi$ Per as they both are binaries with sdO companion and also have been reported similar profile variations. The emission sometimes show rapid variability. The rapid variability is likely to be associated with the changes in the distribution of material within the disk. \cite{maintz2005} reported the phase-locked emission variability of 59 Cyg to have a period of 28.192 days. The appearance of the third peak has been observed for a few other well known classical Be stars like $\phi$ Per, $\zeta$ Tau, $\nu$ Gem, $\chi$ Gem, Pleione.

\cite{stefl2007} reported V/R variations of $\zeta$ Tau, $\nu$ Gem, $\phi$ Per along with others. They reported that a triple-peak profile was observed in $\zeta$ Tau and $\nu$ Gem. They observed that the triple-peak appears only in a particular part of the phase. In the case of $\zeta$ Tau, the triple-peak appeared during the \(V < R\) to \(V > R\) transition, whereas it appeared 
during the \(V > R\) to \(V < R\) transition for $\nu$ Gem. The triple peak and the V/R variation detected in 59 Cyg, is similar to that of $\zeta$ Tau. Thus stars which show  triple-peak profile can be classified into two types:\\ 
\begin{enumerate}
\item \(V > R\) to \(V < R\) transition: Case-I (e.g. $\nu$ Gem)
\item \(V \leq R\) to \(V > R\) transition: Case-II (e.g. $\zeta$ Tau, 59 Cyg)
\end{enumerate} 

The Case-I can be seen as a case of a perturbation moving in the disk, with the same sense of rotation of the disk, and hence a prograde case. \cite{stefl2007} stated that the  triple-peak case can neither be explained by Okazaki's model \citep{okazaki1991} nor by  the rapidly expanding circumstellar ring model of \citep{arias2007}. They conclude that the phase of the m = 1 oscillation and assumed Keplerian rotation are inconsistent with large radial velocity fields in the disks. \cite{stefl2007} mentioned that disks with large eccentricity can precess in a prograde direction and $\nu$ Gem is found to have large eccentricity. In the case of Case-II, the perturbation has to move with the opposite sense of rotation and hence can be termed as the retrograde case. Thus, in 59 Cyg and in $\zeta$ Tau, the appearance of triple-peak and the V/R variation are in the opposite sense and is thus, a retrograde case. This retrograde kind of motion is difficult to explain since the density sub-structure has to move against the sense of rotation of the disk to be seen as the third peak. \cite{stefl2007} also reiterated that the appearance of triple-peak is not related to the binary period. 
59 Cyg, has a very large radial velocity field due to the large H$\alpha$ emission region as reported in our study. We have been able to capture this short lived triple-peak emission in H$\alpha$, which is generally seen in binary systems. We still do not fully understand the triple-peak profile. The triple-peak feature is not seen in many stars and our study shows that all such stars are different from each other. Further observations and continuous monitoring of such stars during different V/R phase can provide valuable inputs to the understanding of processes in the circumstellar disk.

OT Gem underwent a short-term disk loss during our observations. We capture the phase of disk loss and detect a change in the disk radius of the star within three months of observations. During the dissipation of the disk, the H$\alpha$ emitting region shrank, from about 6.9 to 1.7. This might suggest that the disk is dissipated from outside to inside. The outer disk is lost initially and later the feeble emission is still seen due to the inner disk. This star showed a disk loss in a very short time scale unlike the general duration of years to decades. Also, at present from BeSS database \citep{neiner2011}, this star is seen to have emission which means that the star is building the disk again and would be an interesting case to determine the period of such variability. This is the second time that the emission has reappeared after our 2009 observations. We demonstrate that monitoring of these systems are important to understand the way in which the Be star disk is dissipated as well as rejuvenated. 

Our study reveals that short-term monitoring is important to understand different types of variations seen in classical Be stars. The short term variations which are detected in our observations give valuable insights into their circumstellar disk and the physical processes governing the material in the disk. This study also is important, especially since these observations can be performed with moderate telescope equipped with a spectrograph. 

\section{Conclusions}
\label{Sect5}
\begin{enumerate}
\item We have presented the spectroscopic analysis of the two Classical Be stars, 59 Cyg and OT Gem which were observed in 2009.
\item The rotational velocity, $\textit{v}$ sin $\textit{i}$ was calculated using He{\sc i} lines and is found to be $\sim$ 400$\rm km $ $ s^{-1}$ for 59 Cyg and $\sim$ 130$\rm km $ $ s^{-1}$ for OT Gem . The fraction of critical rotation for 59 Cyg is found to be $\sim$ 0.8, suggesting it to be a rapid rotator and $\sim$ 0.3 for OT Gem, indicating it to be a slow rotator or viewed in low inclination.
\item The radius of the circumstellar disk \( R_d/R_*\) using the H$\alpha$ double-peaked emission profile is found to be $\sim$ 10.0, assuming a Keplerian orbit for 59 Cyg. This implies that the H$\alpha$ emission disk is very large for 59 Cyg.
\item Triple-peak is observed for 59 Cyg and the H$\alpha$ profile variation is also detected. We make an attempt to classify stars observed to show triple-peak into two groups. The mechanisms for the formation of triple-peak in both the cases remains a mystery.
\item The EW of the H$\alpha$ emission line profile for OT Gem varied from -4.1 to -0.7~\AA ~in a span of four months.
\item OT Gem underwent a disk loss phenomenon during our observations and lost the outer disk very rapidly but continued to have a feeble emission from the inner disk. This suggests that the disk loss happened from outside to inside during this phase.
\item We conclude and confirm that these two stars show rapid short-term variations in  H$\alpha$ emission and their close
monitoring is important to understand the physical processes in their circumstellar disk.
\end{enumerate}

\normalem
\begin{acknowledgements}
This work was funded by Centre for Research, Christ University, Bangalore as a part of Major Research Project. We thank the support staff at 1.0m telescope, Vainu Bappu Observatory, especially Jayakumar K., for their assistance in obtaining the data used in this paper.
\end{acknowledgements}

\bibliographystyle{raa}
\bibliography{ref1}

\end{document}